\documentclass[traditabstract]{aa}
\usepackage{times}
\usepackage{graphics,graphicx,natbib}

\bibpunct{(}{)}{;}{a}{}{,}

\def\ltsima{$\; \buildrel < \over \sim \;$}
\def\lsim{\lower.5ex\hbox{\ltsima}}
\def\gtsima{$\; \buildrel > \over \sim \;$}
\def\gsim{\lower.5ex\hbox{\gtsima}}

\def\xmm{{\em XMM--Newton}}

\newcommand{\src}{AX \,J1818.8$-$1559}

\begin{document}

\title
{The magnetar candidate AX \,J1818.8$-$1559}
\author{S. Mereghetti\inst{1}, P. Esposito\inst{1},  A. Tiengo\inst{2,1}, D. G\"otz\inst{3}, G.L. Israel\inst{4}, A. De Luca\inst{1}}

\institute{INAF -- Istituto di Astrofisica Spaziale e Fisica
Cosmica Milano, via E.\ Bassini 15, I-20133 Milano, Italy
\and
IUSS -- Istituto Universitario di Studi Superiori, viale Lungo Ticino Sforza 56, I-27100 Pavia, Italy
\and
AIM -- CEA/Irfu/Service d'Astrophysique, Orme des Merisiers, F-91191 Gif-sur-Yvette, France
\and
INAF -- Osservatorio Astronomico di Roma, via Frascati 33, I-00040 Monteporzio Catone, Italy
}

\offprints{S. Mereghetti, sandro@iasf-milano.inaf.it}

\date{Received June 27, 2012 / Accepted August 1, 2012}

\authorrunning{S. Mereghetti et al.}

\titlerunning{}

\abstract { In October 2007 a hard X-ray burst  was detected  by the $INTEGRAL$ satellite from a
direction consistent with the position of \src , an X-ray source at low Galactic latitude
discovered with the \emph{ASCA} satellite in 1996-1999. The short duration (0.8 s) and soft
spectrum (power law photon index of 3.0$\pm$0.2) of the burst in the  20--100 keV range are typical
of Soft Gamma-ray Repeaters and Anomalous X--ray Pulsars. We report on the results of an
observation of \src\ obtained with the \emph{Suzaku} satellite in October 2011. The source
spectrum, a power law with photon index $\sim$1.5, and flux  ($\sim2\times10^{-12}$ erg cm$^{-2}$
s$^{-1}$, 2-10 keV) do not show significant variations with respect to the values derived from
archival data of various satellites ($ROSAT$, \xmm , $Chandra$, $Swift$) obtained from 1993 to
2011. We discuss possible interpretations for \src\ and, based on its association with the
\emph{INTEGRAL} burst, we propose it as  a new member of the small class of magnetar candidates.
}

\maketitle

\section{Introduction}

Although early suggested (e.g. \citealt{pac67}), the relevance of magnetic energy in powering the emission from neutron stars has
been recognized only recently, with the discovery of
anomalous X-ray pulsars (AXPs) and  soft gamma-ray repeaters (SGRs).
These are spinning-down, isolated neutron stars with luminosity in the soft and hard X-rays
typically larger than their rotational energy loss and characterized  by the emission of
powerful bursts and flares.
AXPs and SGRs were historically divided in two classes, but many observations indicate that there are
no substantial differences betwen them (see \citealt{mer08} for  a review).
Even though alternative interpretations have been proposed
\citep{cha00a,alp01,cea06,xu07,hor07,ouy07a,ert09},
the model involving magnetars, i.e. highly magnetized neutron stars, is the one that currently better
explains the properties of AXPs and SGRs \citep{dun92,tho95,tho02}.
According to the magnetar model,  the emission from these sources is ultimately powered by the energy stored in
their strong magnetic fields, with   $B\sim10^{13}$--10$^{14}$ G in the magnetosphere,
and possibly   reaching even larger values in the neutron star interior.

The originally small sample of AXPs/SGRs has  increased significantly in the latest years, mainly
thanks to the discovery of transient sources,
which remained undetected (or unnoticed) until they underwent bright outbursts, often accompanied
by the emission of short bursts \citep{rea11}.
In a few other  cases, new magnetar candidates were found
by thorough investigation of persistent X-ray sources with peculiar properties and/or
possible associations with supernova remnants (e.g. \citealt{gel07,hal10b,hal10c}),
or by the  discovery of a peculiar radio  pulsar  \citep{lev10,and12}.

Many of the new additions to the magnetar family have shown that these objects present a
variety of properties unsuspected from the original members of the class and which connect them
with other classes of isolated neutron stars. The radio detections of XTE\,J1810--197 and
1E\,1547--54 \citep{cam06,cam07c} and the very small spin-down rate of SGR\,0418+5729
\citep{rea10}, implying a dipole field close to that of normal pulsars, are just a few examples.

The X-ray source  \src\ was discovered during a survey of the Galactic plane carried out with \emph{ASCA}
in 1996--1999 \citep{sug01}, but it did not attract particular attention
until  \emph{INTEGRAL} detected a short burst  from its direction in  October 2007.
Here we report  on a detailed re-analysis of the \emph{INTEGRAL} burst and on the results of a long X-ray observation of \src\
carried out with the \emph{Suzaku} satellite in October 2011. We also present  archival X-ray data, that were mostly unpublished,
to provide an exhaustive description of the X-ray properties of \src ,  and discuss its possible
inclusion in the class of magnetar candidates.

\begin{table*}[htbp]
\caption{ Log of the observations. \label{tab-obs}}
\begin{center}
\begin{tabular}{lclcl}
\hline
 Satellite          &  Obs.Id   &  date   &  Duration (ks)  & Notes  \\
\hline
\hline
\emph{ROSAT }             &   rp50311n00       &   1993 September 12     & 6.7  &  PSPC \\
\emph{XMM-Newton}         &   0152834501       &  2003 March 28           &  10.0      &  EPIC \\
\emph{Swift}              &   00020057001      &   2007 October 19        &   10.4         &  XRT \\
\emph{Chandra   }         &    7617            &  2007 October 21-22      &    5.1          & ACIS-S  TE \\ 
\emph{Chandra }           &    7616           &  2007 October 22         &   39.9        & ACIS-S  CC \\ 
\emph{Chandra}            &    9609             &   2008 February 19       &    1.6         & ACIS   \\
\emph{Chandra}            &    9610             &    2008 September 25     &   1.2          & HRC-I, 12$'$ off-axis \\ 
\emph{Suzaku}             &    406074010        &   2011 October 15-18     & 80  & XIS \\
\emph{Swift }             &    00044118001    &  2012 March 4     &      0.5      &  XRT \\
\hline
\end{tabular}
\end{center}

\end{table*}

\section{The burst of 2007 October 17}

\begin{figure}
 \includegraphics[width=80mm,angle=-90]{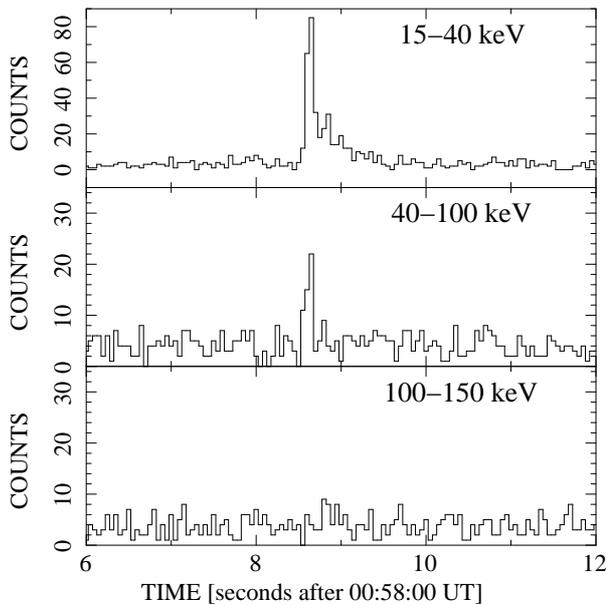}
\caption{\footnotesize {Light curves of the  2007 October 17 burst measured
with ISGRI in three energy ranges. The bin size is 50 ms. }
 \label{fig-lcburst}}
 \end{figure}

A short burst at low Galactic latitude (b=--0.26),  was discovered and localized in real
time with the \emph{INTEGRAL} Burst Alert System (IBAS; \citealt{mer03}) on 2007 October 17. The
burst, reported as GRB 071017 \citep{mer07gcn}, was found in the data of the \emph{ISGRI} detector
\citep{leb03}, which is the low-energy imager of the \emph{IBIS} coded-mask instrument on board
\emph{INTEGRAL} \citep{ube03}. The burst light curves in different energy ranges
(Fig.~\ref{fig-lcburst}) show that, unlike the typical gamma-ray bursts (GRBs) observed by  IBIS
\citep{via09}, this event was not detected above 100 keV. The 15--40 keV light curve is
characterized by a single pulse lasting $\sim$0.8 s, with a fast rise time starting at 00:58:08.55
UT followed by a slower decay.

\begin{figure}
 \includegraphics[width=65mm,angle=-90]{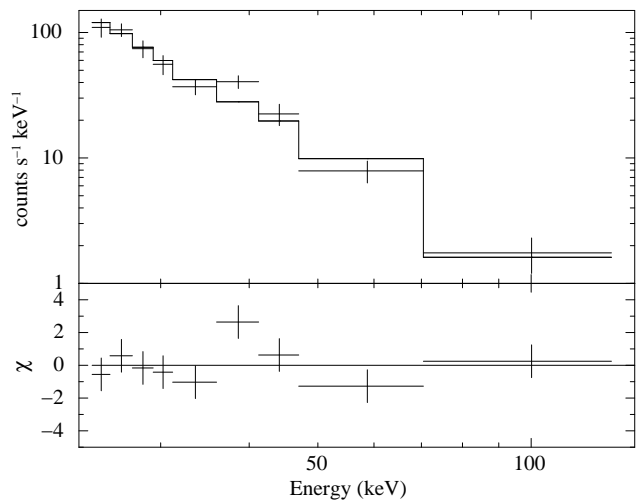}
\caption{\footnotesize {ISGRI spectrum of the  2007 October 17 burst. Top panel:
data and best fit power law model. Bottom panel: residuals in units of sigma.  }
\label{fig-spectrum-burst}}
 \end{figure}

We performed a spectral analysis by extracting the ISGRI data of the first 0.2 s, corresponding to the
brightest part of the burst.
The results confirm that, as suggested by the light curves in different energyt bands,
the burst was rather soft. Its spectrum    is
equally well fit by a power law with  photon index  $3.0\pm0.2$ or by  a thermal
bremsstrahlung   with temperature kT=25$\pm$3 keV   (Fig.~\ref{fig-spectrum-burst}).
In both cases the average flux in the 20-100 keV energy range is $\sim2\times10^{-7}$ erg  cm$^{-2}$ s$^{-1}$.

The burst localization that was obtained in real time by the automatic IBAS software had an error of 2.9$'$ and was consistent
with the position of \src . We reanalyzed    the ISGRI data using the most recent versions of the satellite
attitude information and  imaging software. Using the data  in the 15-50 keV, where the signal to noise
is maximum, we obtained a 11.5$\sigma$ detection of the burst at  the coordinates
R.A.=274.731, Dec.=$-–$15.993 (J2000), with an uncertainty of 1.8$'$ (90\% confidence level
radius). The Corresponding galactic coordinates are  l=15.04, b=--0.26. The
\textit{INTEGRAL} error region is indicated by the   circle in Fig.~\ref{fig-imaEPIC}.

\section{\xmm\  archival data}

\begin{figure}
 \includegraphics[width=85mm]{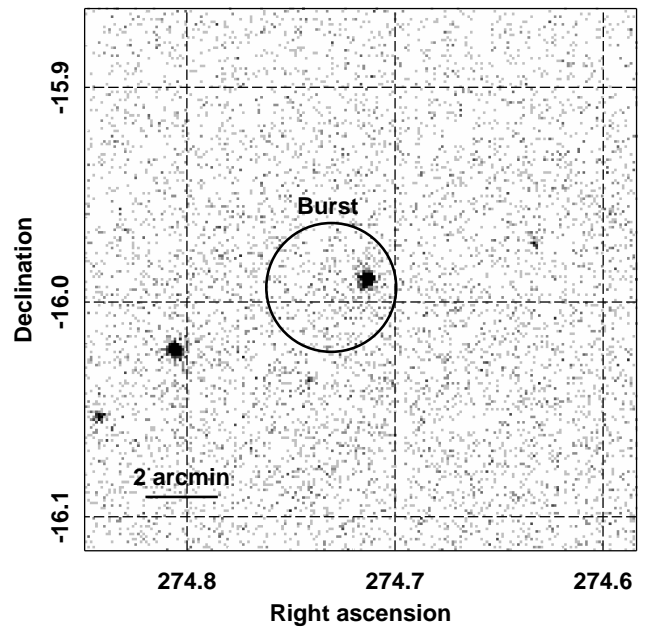}
\caption{\footnotesize {Image of the region of \src\ in the 2-10 keV energy
range obtained with the \xmm\  EPIC instrument. The  circle (radius 1.8$'$)
indicates the error region of the \emph{INTEGRAL} burst. }
\label{fig-imaEPIC}}
 \end{figure}

The region of sky in the direction of GRB 071017 was observed for about 10 ks on 2003 March 28,  during  a survey
of the Galactic plane carried out with   the \xmm\ satellite.
Using the   Science Analysis Software (SAS V11)
we reprocessed the data of the EPIC instrument (0.2--12 keV), which consists
of one pn \citep{tur01} and two MOS \citep{str01} cameras.  A few X-ray sources are detected in the EPIC   images, but only
one, at an off-axis angle of 4$'$, lies in the error circle of the \emph{INTEGRAL} burst (Fig. \ref{fig-imaEPIC}).
Its  coordinates are $\rm R.A.=18^{h}18^{m}51\fs5$, $\rm Dec.=-15\degr59'23''$  (J2000), with an uncertainty of  2$''$.
Based on the positional coincidence, as well as on the consistent   spectral properties (see next paragraph),
we identify the \xmm\ source with    \src\ \citep{sug01}.

No other sources are detected with \xmm\ inside the error circle of the burst, down to a flux limit of a few 10$^{-14}$
erg cm$^{-2}$ s$^{-1}$.  A few other sources are visible in the EPIC image. A faint source is located 1$'$ from the
southern border of the error circle, at $\rm R.A.=18^{h}18^{m}57\fs9$,  Dec.=--16$^{\circ}$ 02$'$ 08$''$ (J2000).
Its  flux is about a factor 20 smaller than that of \src .
A brighter source, at  $\rm R.A.=18^{h}19^{m}13\fs6$,  Dec.=--16$^{\circ}$ 01$'$ 20$''$ (J2000), can be
identified with the \emph{ASCA} source  AX J181915--1527  \citep{sug01}.

To study the   spectrum of \src , we extracted the pn and MOS source counts from a circle of 40$''$
radius and the background spectra from source free regions of the same CCD as the source. After
rebinning to have at least 30 counts in each spectral bin, we fitted simultaneously the pn and two
MOS spectra. A good fit was obtained  with a power law with photon index $\Gamma=1.2\pm0.2$,
absorption $N_{\rm H} = (3.6\pm0.5)\times10^{22}$ cm$^{-2}$, and absorbed\footnote{We use the
\texttt{PHABS} absorption model in XSPEC with abundances from \cite{and89}.} flux
$1.4\times10^{-12}$ erg cm$^{-2}$ s$^{-1}$  in the 2--10 keV energy range. These results are
consistent   with those we obtained in  a quick look analysis of the same data performed soon after
the \emph{INTEGRAL} GRB detection \citep{tie07atel}. The source distance is unknown, but
considering its location in the direction of the Galactic center region and the large absorption,
we assume in the following a reference value of 10 kpc.
The 0.5--10 keV luminosity, corrected for the absorption, is $\sim$$2.4\times10^{34}$ erg
s$^{-1}$. An equally good fit is obtained with a blackbody of temperature $kT_{BB}=1.9\pm0.1$ keV
and emission radius $R_{\rm BB}=0.12\pm0.05$ km.
In this case the absorption is $N_{\rm H} = (1.6\pm0.3)\times10^{22}$ cm$^{-2}$. For
comparison, the total Galactic absorption in this direction is 1.2$\times10^{22}$ cm$^{-2}$
\citep{kal05}. All the spectral results are summarized in Table \ref{tab-spectra}. The values
obtained with the power law model are very similar to those measured with \emph{ASCA} in 1996--1999
\citep{sug01}.

We searched for pulsations in the EPIC data with  Fast Fourier Transform (FFT) and Rayleigh
test methods without finding any statistically significant signal. However, this result is not
particularly constraining due to the relatively small number of counts.
The 3$\sigma$ upper limits on the pulsed fraction, A, in the 1-10 keV energy band are of 50\% for
periods in the 0.3-1000 s range and of $\sim$50-70\% between 145 and 300 ms\footnote{Here and in
the following we assume a sinusoidal modulation of the flux, F(t)=F$_o$ [1+Asin($\Omega$t)]
and define the pulsed fraction as the sinusoid amplitude A. All the upper limits on A are
computed as described in \citet{isr96}}..

 \begin{figure}
 \includegraphics[width=88mm]{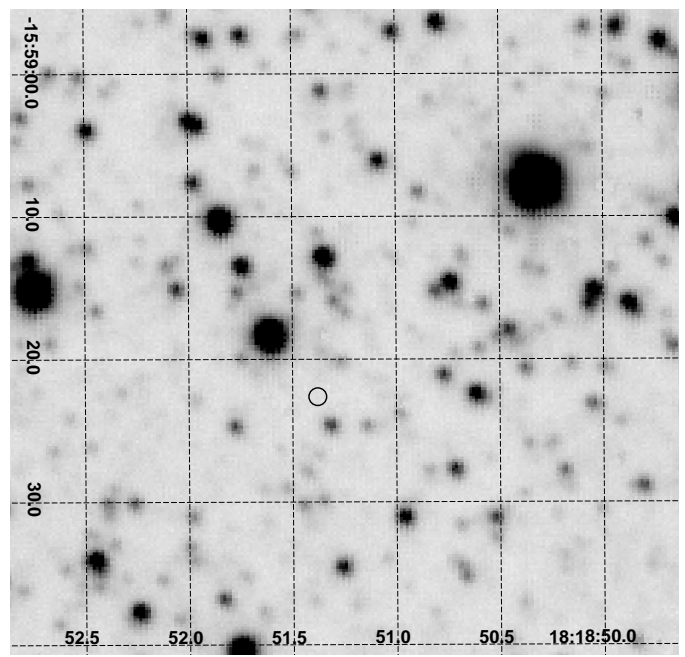}
\caption{\footnotesize {Near infrared (K-band) image of the region of \src\ from the UKIDSS. The  circle with radius
0.6 arcsec indicates the source
position derived with \emph{Chandra}.  }
\label{fig-optical}}
 \end{figure}

\section{\emph{Suzaku} observation}

We performed  a \emph{Suzaku} observation of  \src\   from 2011 October 15 at 13:17 UT to October
18 at 03:32 UT. This provided about 80 ks of exposure in the XIS instrument. The XIS consists of
four telescopes with a spatial resolution of about 2$'$ coupled to four CCD cameras operating in
the 0.2--12 keV energy range   \citep{koy07}. At the time of our observation only the two front
illuminated XIS0 and XIS3, and the back-illuminated XIS1 were  operating. The three detectors were
used in 1/8 window mode, which gives a time resolution of 1 s at the expense of a reduced field of
view.

For the spectral analysis we screened the data using standard criteria. In particular, we
considered only events with GRADE = 0,2--4,6 and excluded time intervals within $\Delta$T = 436 s
from the South Atlantic Anomaly (SAA) and pointed at elevation angles smaller than  5\degr\ or
20\degr\ above the night or day Earth rim, respectively. Source and background spectra were
extracted from circles of 1.1$'$ radius and rebinned so as to have a minimum of 20 counts per
spectral bin. Slightly less stringent criteria were used for the timing analysis ($\Delta$T = 60 s,
10\degr\ day Earth minimum elevation) in order to increase the exposure time, still keeping a
reasonably low background.

We obtained a single spectrum by merging the XIS 0 and 3 (the two front-illuminated CCDs) data and
fitted it simultaneously with the XIS1 spectrum. All the fits were done in the 1--10 keV range.
Both a power law and a blackbody model gave acceptable results, with the best fit  parameters
reported in Table \ref{tab-spectra}. These values are similar to those derived in the \xmm\
observation.

We searched for periodicities in the 2--1000 s range  by an FFT analysis on  the merged
event lists from the three instruments.
No periodicities were found, even trying different
energy selections. The 3$\sigma$ upper limit  on the 1-10 keV pulsed fraction is $\sim$30-40\% between
2 and 40 s, and $\sim$25-30\% above 40 s.

\begin{table*}[htbp]
\caption{Spectral results. \label{tab-spectra}  }
\begin{center}
\begin{tabular}{lcccc}
\hline
                                              &  \xmm\  & \emph{Chandra} &  \emph{Suzaku}    \\
                                              & 2003     & 2007          &  2011   \\
 \hline
\hline
$\Gamma$                                     &      1.17$\pm$0.17        & $1.56\pm0.22$          & $1.53\pm0.14$  \\
N$_H$  (10$^{22}$ cm$^{-2}$)                 &      3.6$\pm$0.5          &  $4.5_{-0.5}^{+0.6}$   &    $6.3_{-0.7}^{+0.8}$   \\
F$^a$ (10$^{-12}$ erg cm$^{-2}$  s$^{-1}$)   &    1.37$\pm$0.07          & $1.02_{-0.06}^{+0.07}$ &   $1.04\pm0.03$\\
F$^b$ (10$^{-12}$ erg cm$^{-2}$  s$^{-1}$)   &    2.11$_{-0.13}^{+0.17}$ &  $2.0_{-0.2}^{+0.3}$   &    $2.19_{-0.19}^{+0.16}$\\
$\chi^2$/dof, dof                           &    0.88, 79                 & 1.10, 91               &  1.07, 155 \\
\hline
kT$_{BB}$  (keV)                             &    1.87$\pm$0.12        & $1.42^{+0.11}_{-0.09}$   &    1.84$_{-0.08}^{+0.09}$    \\
N$_H$  (10$^{22}$ cm$^{-2}$)                 &    1.6$\pm$0.3          & $2.5\pm0.3$              & 2.5$_{-0.4}^{+0.5}$\\
R$_{BB}^{c}$  (km)                           &   0.12$\pm$0.05         & $0.16\pm0.02$              &  $0.11\pm0.01$\\
F$^a$   (10$^{-12}$ erg cm$^{-2}$  s$^{-1}$)  &    1.26$\pm$0.07      &  $8.4_{-0.5}^{+0.6}$      & $1.00\pm0.03$ \\
F$^b$   (10$^{-12}$ erg cm$^{-2}$  s$^{-1}$)  &   1.45$\pm$0.06        & $1.09\pm0.05$             &    $1.21\pm0.04$\\\
$\chi^2$/dof, dof                           &      0.92, 79           & 1.12, 91                  &    1.07, 155\\
\hline
\end{tabular}
\end{center}
\begin{list}{}{}
\item  Quoted errors are at 1$\sigma$ confidence level.
\item $^a$     2--10 keV    flux.
\item $^b$    0.5--10 keV flux  corrected for absorption.
\item $^d$ For $d=10$ kpc.
\end{list}
\end{table*}

\section{Other archival data}

\subsection{Swift}

A short observation (2 ks) with Swift/XRT  was performed on  2007 October 19,  about 2 days after the \emph{INTEGRAL}
detection of the  burst \citep{eva07gcn}.
A source with count rate $0.012\pm0.002$ counts s$^{-1}$ was  detected at the coordinates
$\rm R.A.=18^h18^m51\fs49$, $\rm Dec.=-15\degr59'20\farcs07$ (J2000; $4\farcs1$ error radius, 90\% confidence)
consistent with those of \src . The small number of counts is not sufficient to
perform a spectral analysis. Assuming the parameters of the \xmm\ power-law fit, we derived
a 2--10 keV observed flux of   $\sim$$1.5\times10^{-12}$ erg cm$^{-2}$ s$^{-1}$.

This region of sky was observed again with the Swift/XRT  on 2012 March 04, but no source was
detected within or close to the error region of the burst. The short exposure time (0.5 ks) and
off-axis position result in a rather shallow upper limit on the count rate of 0.027 counts s$^{-1}$
(3 $\sigma$ c.l.),  which is a factor $\sim2$ above the flux level measured in October
2007.

\subsection{Chandra}

The \emph{Chandra} observations of the field of \src\ are listed in Table \ref{tab-obs}\footnote{Another observation was performed in August 2009 (obs. ID 10503),
but the position of \src\ fell in the gap between the ACIS-S CCDs}.
The  2007 observations   were done with the ACIS-S instrument as a Target of Opportunity   5 days after the  \emph{INTEGRAL} burst
detection. They consist of 5 ks  of imaging data, with time resolution of 3.2 s,  and
a longer exposure in continuous clocking (CC) mode, which provides a higher time resolution (2.85 ms), but imaging
only along a single direction. \src\ is detected with a significance above 10$\sigma$ in the imaging data,
at coordinates  $\rm R.A.=18^{h}18^{m}51\fs38$ ($\pm0\fs06$),  $\rm Dec.=-15\degr59'22\farcs62$  ($\pm0\farcs05$) (J2000); the quoted 1$\sigma$ uncertainties are
statistical only, and much smaller than the \emph{Chandra} absolute astrometric accuracy, $0\farcs6$  (radial, 90\% c.l.).
We show in Fig. \ref{fig-optical} the \emph{Chandra} error circle overlaid on the near infrared (K-band) image from the  Galactic Plane Survey, collected as a part of the
United Kingdom Infrared Deep Sky Survey (UKIDSS, \citet{luc08}).
No  counterparts are visible  down to  the  limit K$\sim$17--18.

We extracted the spectra from the 2007 observations as follows.
For the CC data, the source photons were accumulated within a $5 \times 5$ pixels region centred on \src\ (one ACIS-S pixel corresponds to 0.492 arcsec);
the background events were extracted from two symmetric (with respect to the source) regions along the image strip.
For the imaging data, the source counts were selected in a $\sim$5-px radius circular region and the background in an annulus with radii of 20 and 30 pixels.
After checking that they gave consistent results, the two spectra were fitted together yielding
the best fit parameters reported in Table \ref{tab-spectra}.

\src\ was also observed two further times in 2008. In the first data set (2008 February) the source
is detected at a significance level of 5$\sigma$, but the small statistics do not allow us to carry
out a detailed analysis. We just note that the 0.3--7 keV count rate, $0.026\pm0.004$ counts
s$^{-1}$, is in line with that observed in the 2007 ACIS-S imaging observation,  $0.024\pm0.002$
counts s$^{-1}$. In the second data set (2008 September), a short HRC-I exposure, the source is not
detected. The 3$\sigma$ upper limit on the count rate is 0.026 counts s$^{-1}$ which,
adopting the \emph{Chandra} power-law model of Table \ref{tab-spectra}, translates into an upper
limit on the observed 2-10 keV flux of $\sim$$5\times10^{-12}$ erg cm$^{-2}$ s$^{-1}$.

The most sensitive data for timing analysis are provided by the 2007 data in  CC mode, which
yielded $\sim$1400   source photons. By an FFT  analysis  of these data, we could set a
3$\sigma$ upper limit of $\sim$40\% on the 1-10 keV pulsed fraction for periods in the range
0.01--1000 s, and $\sim$40--60\% for periods between 5 and 10 ms.

\subsection{ROSAT}

The position of \src\ was serendipitously covered by a ROSAT observation on 11-12 September 1993,
falling at an off-axis angle of $\sim$35$'$ in the PSPC instrument. From these data, providing an
exposure of $\sim$7 ks we derived a 3$\sigma$ upper limit of $\sim$0.01 counts s$^{-1}$ on the
source count rate in  the 0.1-2.4 keV energy range. Assuming the same spectral shape of the \xmm\
observation, this corresponds to an unabsorbed flux of the order of 2$\times10^{-11}$ erg cm$^{-2}$
s$^{-1}$ in the 0.5-10 keV range.

\section{Discussion}

A subclass  of GRBs, the so called X-ray Flashes \citep{hei01}, can  have spectra as soft as
observed in GRB 071017. However, in the class of short GRBs with duration less than 2 s, such soft
spectra are never observed. On the other hand, the hard X-ray   (E$>$20 keV)  spectra of short
bursts from SGRs are usually well described by thermal bremsstrahlung models with temperature
$\sim$20-40 keV \citep{apt01,goe04}, as observed in GRB 071017. Assuming a distance of 10 kpc, the
peak luminosity of the \emph{INTEGRAL} burst is of the order of $\sim5\times10^{39}$ erg s$^{-1}$,
which is a reasonable value for a Galactic SGR. Thus, although we cannot completely exclude that
GRB 071017 was a GRB seen through the Galaxy, we believe that, in view  of its location  in
the Galactic plane
and   spectral
properties, an explanation in terms of a Galactic AXP/SGR is much more likely.

The chance probability of finding an X-ray source of a given brightness inside the 10 arcmin$^2$ \emph{INTEGRAL} error circle can be estimated
using the LogN-LogS relations derived by \citet{sug01}   from the survey of the Galactic plane in which \src\ was discovered.
Including the small contribution from extragalactic sources seen through the Galactic plane, these relations give a surface density of 1.2 deg$^{-2}$,
for sources with flux equal or greater than that of \src .
This leads to a   probability of $\sim3\times10^{-3}$ to find \src\ in the \emph{INTEGRAL} error region by chance,
which indicates its very likely association with the burst.

The X-ray spectrum of \src\ is harder than that of typical SGRs and AXPs, which, when fitted with a
single power law, give  photon index in the range $\sim2-4$ \citep{mer02b,kas10}. Harder
spectra have been observed in transient magnetars during the initial phases of their outbursts, but
our analysis of X-ray observations of \src\ obtained from 2003 to 2012, as well as the earlier
\emph{ASCA} observations \citep{sug01}, do not show evidence for a transient behavior.  Indeed all
the measured fluxes, and the upper limits, indicate  a steady luminosity of    $2\times10^{34}$
(d/[10 kpc])$^2$ erg s$^{-1}$, consistent with that observed in persistent magnetars. The lack of
detection of pulsations is not particularly  constraining, considering that many of the AXPs and
SGRs have pulsed fractions smaller than the upper limits we could set on \src\
\citep{mer02b}.

Only a single burst has been observed from the direction of \src , despite this region of sky in
the Galactic bulge having been repeatedly observed by high-energy satellites. This contrasts with
the abundance of bursts, often grouped in time intervals of spasmodic activity, observed in the
most active members of the SGR family (e.g., SGR 1806--20 \citep{gog00}, SGR 1900$+$14
\citep{isr08}, SGR 1627--41 \citep{esp08},  or 1E 1547.0--5408 \citep{mer09}). The behavior of
\src\ is instead similar to that of a few magnetars from which bursts were detected only
sporadically, such as for example 1E 1048.1--5937 \citep{gav02},  SGR 0418$+$5729
\citep{van10},  SGR 1833--0832 \citep{gog10}, or 4U 0142$+$61 \citep{gav11}.

We finally note that the available observations do not favour alternative explanations for the
nature of \src . Coronal emission from a star can be ruled out by the lack of a bright optical
counterpart,  the hard spectrum, and the large absorption, inconsistent with the small distance
that would be required in this hypothesis. The X-ray to optical flux ratio excludes the possibility
of a high mass X-ray binary. A low mass X-ray binary or an AGN cannot be excluded, but they are not
supported by the hard spectrum and lack of long term variability, respectively.

\section{Conclusion}

A new observation of \src\ obtained with the Suzaku satellite, as well as a comprehensive analysis
of archival data from different missions could not establish the real nature of this source, which
is most likely associated with a SGR-like burst observed with \emph{INTEGRAL} in October 2007. The
X-ray properties of \src\  are consistent  with those observed in AXPs and SGRs, and  we propose it
as a likely new member of this class of sources deserving further study.

\begin{acknowledgements}

This work was partially supported with
contributions from the agreements ASI-INAF I/009/10/0 and I/032/10/0.
This research has made use of data obtained from the \emph{Suzaku} satellite, a collaborative mission between
the space agencies of Japan (JAXA) and the USA (NASA).
We also made use of data and software provided by the NASA/GFSC's HEASARC, the \emph{Chandra}
X-ray Centre and the ESA's \xmm\ Science Archive.
We thank the referee, V. Kaspi, for her useful comments.

\end{acknowledgements}

\bibliographystyle{aa}
\bibliography{axpsgr6}

\begin{thebibliography}{49}
\expandafter\ifx\csname natexlab\endcsname\relax\def\natexlab#1{#1}\fi

\bibitem[{{Alpar}(2001)}]{alp01}
{Alpar}, M.~A. 2001, \apj, 554, 1245

\bibitem[{{Anders} \& {Grevesse}(1989)}]{and89}
{Anders}, E. \& {Grevesse}, N. 1989, \gca, 53, 197

\bibitem[{{Anderson} {et~al.}(2012){Anderson}, {Gaensler}, {Slane}, {Rea},
  {Kaplan}, {Posselt}, {Levin}, {Johnston}, {Murray}, {Brogan}, {Bailes},
  {Bates}, {Benjamin}, {Bhat}, {Burgay}, {Burke-Spolaor}, {Chakrabarty},
  {D'Amico}, {Drake}, {Esposito}, {Grindlay}, {Hong}, {Israel}, {Keith},
  {Kramer}, {Lazio}, {Lee}, {Mauerhan}, {Milia}, {Possenti}, {Stappers}, \&
  {Steeghs}}]{and12}
{Anderson}, G.~E., {Gaensler}, B.~M., {Slane}, P.~O., {et~al.} 2012, \apj, 751,
  53

\bibitem[{{Aptekar} {et~al.}(2001){Aptekar}, {Frederiks}, {Golenetskii},
  {Il'inskii}, {Mazets}, {Pal'shin}, {Butterworth}, \& {Cline}}]{apt01}
{Aptekar}, R.~L., {Frederiks}, D.~D., {Golenetskii}, S.~V., {et~al.} 2001,
  \apjs, 137, 227

\bibitem[{{Camilo} {et~al.}(2007){Camilo}, {Ransom}, {Halpern}, \&
  {Reynolds}}]{cam07c}
{Camilo}, F., {Ransom}, S.~M., {Halpern}, J.~P., \& {Reynolds}, J. 2007, \apjl,
  666, L93

\bibitem[{{Camilo} {et~al.}(2006){Camilo}, {Ransom}, {Halpern}, {Reynolds},
  {Helfand}, {Zimmerman}, \& {Sarkissian}}]{cam06}
{Camilo}, F., {Ransom}, S.~M., {Halpern}, J.~P., {et~al.} 2006, \nat, 442, 892

\bibitem[{{Cea}(2006)}]{cea06}
{Cea}, P. 2006, \aap, 450, 199

\bibitem[{{Chatterjee} {et~al.}(2000){Chatterjee}, {Hernquist}, \&
  {Narayan}}]{cha00a}
{Chatterjee}, P., {Hernquist}, L., \& {Narayan}, R. 2000, \apj, 534, 373

\bibitem[{{Duncan} \& {Thompson}(1992)}]{dun92}
{Duncan}, R.~C. \& {Thompson}, C. 1992, \apjl, 392, L9

\bibitem[{{Ertan} {et~al.}(2009){Ertan}, {Ek{\c s}i}, {Erkut}, \&
  {Alpar}}]{ert09}
{Ertan}, {\"U}., {Ek{\c s}i}, K.~Y., {Erkut}, M.~H., \& {Alpar}, M.~A. 2009,
  \apj, 702, 1309

\bibitem[{{Esposito} {et~al.}(2008){Esposito}, {Israel}, {Zane}, {Senziani},
  {Starling}, {Rea}, {Palmer}, {Gehrels}, {Tiengo}, {de Luca}, {G{\"o}tz},
  {Mereghetti}, {Romano}, {Sakamoto}, {Barthelmy}, {Stella}, {Turolla},
  {Feroci}, \& {Mangano}}]{esp08}
{Esposito}, P., {Israel}, G.~L., {Zane}, S., {et~al.} 2008, \mnras, 390, L34

\bibitem[{{Evans} {et~al.}(2007){Evans}, {Starling}, {O'Brien}, {Israel},
  {Beardmore}, {Page}, {Osborne}, {Cummings}, \& {Gehrels}}]{eva07gcn}
{Evans}, P.~A., {Starling}, R.~L.~C., {O'Brien}, P.~T., {et~al.} 2007, GRB
  Coordinates Network, 6942, 1

\bibitem[{{Gavriil} {et~al.}(2011){Gavriil}, {Dib}, \& {Kaspi}}]{gav11}
{Gavriil}, F.~P., {Dib}, R., \& {Kaspi}, V.~M. 2011, \apj, 736, 138

\bibitem[{{Gavriil} {et~al.}(2002){Gavriil}, {Kaspi}, \& {Woods}}]{gav02}
{Gavriil}, F.~P., {Kaspi}, V.~M., \& {Woods}, P.~M. 2002, \nat, 419, 142

\bibitem[{{Gelfand} \& {Gaensler}(2007)}]{gel07}
{Gelfand}, J.~D. \& {Gaensler}, B.~M. 2007, \apj, 667, 1111

\bibitem[{{G{\"o}tz} {et~al.}(2004){G{\"o}tz}, {Mereghetti}, {Mirabel}, \&
  {Hurley}}]{goe04}
{G{\"o}tz}, D., {Mereghetti}, S., {Mirabel}, I.~F., \& {Hurley}, K. 2004, \aap,
  417, L45

\bibitem[{{G{\"o}{\u g}{\"u}{\c s}} {et~al.}(2010){G{\"o}{\u g}{\"u}{\c s}},
  {Cusumano}, {Levan}, {Kouveliotou}, {Sakamoto}, {Barthelmy}, {Campana},
  {Kaneko}, {Stappers}, {de Ugarte Postigo}, {Strohmayer}, {Palmer}, {Gelbord},
  {Burrows}, {van der Horst}, {Mu{\~n}oz-Darias}, {Gehrels}, {Hessels},
  {Kamble}, {Wachter}, {Wiersema}, {Wijers}, \& {Woods}}]{gog10}
{G{\"o}{\u g}{\"u}{\c s}}, E., {Cusumano}, G., {Levan}, A.~J., {et~al.} 2010,
  \apj, 718, 331

\bibitem[{{G{\"o}{\u g}{\"u}{\c s}} {et~al.}(2000){G{\"o}{\u g}{\"u}{\c s}},
  {Woods}, {Kouveliotou}, {van Paradijs}, {Briggs}, {Duncan}, \&
  {Thompson}}]{gog00}
{G{\"o}{\u g}{\"u}{\c s}}, E., {Woods}, P.~M., {Kouveliotou}, C., {et~al.}
  2000, \apjl, 532, L121

\bibitem[{{Halpern} \& {Gotthelf}(2010{\natexlab{a}})}]{hal10c}
{Halpern}, J.~P. \& {Gotthelf}, E.~V. 2010{\natexlab{a}}, \apj, 725, 1384

\bibitem[{{Halpern} \& {Gotthelf}(2010{\natexlab{b}})}]{hal10b}
---. 2010{\natexlab{b}}, \apj, 710, 941

\bibitem[{{Heise} {et~al.}(2001){Heise}, {Zand}, {Kippen}, \& {Woods}}]{hei01}
{Heise}, J., {Zand}, J.~I., {Kippen}, R.~M., \& {Woods}, P.~M. 2001, in
  Gamma-ray Bursts in the Afterglow Era, ed. E.~{Costa}, F.~{Frontera}, \&
  J.~{Hjorth}, 16

\bibitem[{{Horvath}(2007)}]{hor07}
{Horvath}, J.~E. 2007, \apss, 308, 431

\bibitem[{{Israel} {et~al.}(2008){Israel}, {Romano}, {Mangano}, {Dall'Osso},
  {Chincarini}, {Stella}, {Campana}, {Belloni}, {Tagliaferri}, {Blustin},
  {Sakamoto}, {Hurley}, {Zane}, {Moretti}, {Palmer}, {Guidorzi}, {Burrows},
  {Gehrels}, \& {Krimm}}]{isr08}
{Israel}, G.~L., {Romano}, P., {Mangano}, V., {et~al.} 2008, \apj, 685, 1114

\bibitem[{{Israel} \& {Stella}(1996)}]{isr96}
{Israel}, G.~L. \& {Stella}, L. 1996, \apj, 468, 369

\bibitem[{{Kalberla} {et~al.}(2005){Kalberla}, {Burton}, {Hartmann}, {Arnal},
  {Bajaja}, {Morras}, \& {P{\"o}ppel}}]{kal05}
{Kalberla}, P.~M.~W., {Burton}, W.~B., {Hartmann}, D., {et~al.} 2005, \aap,
  440, 775

\bibitem[{{Kaspi} \& {Boydstun}(2010)}]{kas10}
{Kaspi}, V.~M. \& {Boydstun}, K. 2010, \apjl, 710, L115

\bibitem[{{Koyama} {et~al.}(2007){Koyama}, {Tsunemi}, {Dotani}, {Bautz},
  {Hayashida}, {Tsuru}, {Matsumoto}, {Ogawara}, {Ricker}, {Doty}, {Kissel},
  {Foster}, {Nakajima}, {Yamaguchi}, {Mori}, {Sakano}, {Hamaguchi},
  {Nishiuchi}, {Miyata}, {Torii}, {Namiki}, {Katsuda}, {Matsuura}, {Miyauchi},
  {Anabuki}, {Tawa}, {Ozaki}, {Murakami}, {Maeda}, {Ichikawa}, {Prigozhin},
  {Boughan}, {Lamarr}, {Miller}, {Burke}, {Gregory}, {Pillsbury}, {Bamba},
  {Hiraga}, {Senda}, {Katayama}, {Kitamoto}, {Tsujimoto}, {Kohmura}, {Tsuboi},
  \& {Awaki}}]{koy07}
{Koyama}, K., {Tsunemi}, H., {Dotani}, T., {et~al.} 2007, \pasj, 59, 23

\bibitem[{{Lebrun} {et~al.}(2003){Lebrun}, {Leray}, {Lavocat}, {Cr{\'e}tolle},
  {Arqu{\`e}s}, {Blondel}, {Bonnin}, {Bou{\`e}re}, {Cara}, {Chaleil}, {Daly},
  {Desages}, {Dzitko}, {Horeau}, {Laurent}, {Limousin}, {Mathy}, {Mauguen},
  {Meignier}, {Molini{\'e}}, {Poindron}, {Rouger}, {Sauvageon}, \&
  {Tourrette}}]{leb03}
{Lebrun}, F., {Leray}, J.~P., {Lavocat}, P., {et~al.} 2003, \aap, 411, L141

\bibitem[{{Levin} {et~al.}(2010){Levin}, {Bailes}, {Bates}, {Bhat}, {Burgay},
  {Burke-Spolaor}, {D'Amico}, {Johnston}, {Keith}, {Kramer}, {Milia},
  {Possenti}, {Rea}, {Stappers}, \& {van Straten}}]{lev10}
{Levin}, L., {Bailes}, M., {Bates}, S., {et~al.} 2010, \apjl, 721, L33

\bibitem[{{Lucas} {et~al.}(2008){Lucas}, {Hoare}, {Longmore}, {Schr{\"o}der},
  {Davis}, {Adamson}, {Bandyopadhyay}, {de Grijs}, {Smith}, {Gosling},
  {Mitchison}, {G{\'a}sp{\'a}r}, {Coe}, {Tamura}, {Parker}, {Irwin}, {Hambly},
  {Bryant}, {Collins}, {Cross}, {Evans}, {Gonzalez-Solares}, {Hodgkin},
  {Lewis}, {Read}, {Riello}, {Sutorius}, {Lawrence}, {Drew}, {Dye}, \&
  {Thompson}}]{luc08}
{Lucas}, P.~W., {Hoare}, M.~G., {Longmore}, A., {et~al.} 2008, \mnras, 391, 136

\bibitem[{{Mereghetti}(2008)}]{mer08}
{Mereghetti}, S. 2008, \aapr, 15, 225

\bibitem[{{Mereghetti} {et~al.}(2002){Mereghetti}, {Chiarlone}, {Israel}, \&
  {Stella}}]{mer02b}
{Mereghetti}, S., {Chiarlone}, L., {Israel}, G.~L., \& {Stella}, L. 2002, in
  Neutron Stars, Pulsars, and Supernova Remnants, ed. W.~{Becker}, H.~{Lesch},
  \& J.~{Tr{\"u}mper}, 29--+

\bibitem[{{Mereghetti} {et~al.}(2003){Mereghetti}, {G{\"o}tz}, {Borkowski},
  {Walter}, \& {Pedersen}}]{mer03}
{Mereghetti}, S., {G{\"o}tz}, D., {Borkowski}, J., {Walter}, R., \& {Pedersen},
  H. 2003, \aap, 411, L291

\bibitem[{{Mereghetti} {et~al.}(2009){Mereghetti}, {G{\"o}tz},
  {Weidenspointner}, {von Kienlin}, {Esposito}, {Tiengo}, {Vianello}, {Israel},
  {Stella}, {Turolla}, {Rea}, \& {Zane}}]{mer09}
{Mereghetti}, S., {G{\"o}tz}, D., {Weidenspointner}, G., {et~al.} 2009, \apjl,
  696, L74

\bibitem[{{Mereghetti} {et~al.}(2007){Mereghetti}, {Paizis}, {Gotz}, {Petry},
  {Shaw}, {Beck}, \& {Borkowski}}]{mer07gcn}
{Mereghetti}, S., {Paizis}, A., {Gotz}, D., {et~al.} 2007, GRB Coordinates
  Network, 6927, 1

\bibitem[{{Ouyed} {et~al.}(2007){Ouyed}, {Leahy}, \& {Niebergal}}]{ouy07a}
{Ouyed}, R., {Leahy}, D., \& {Niebergal}, B. 2007, \aap, 473, 357

\bibitem[{{Pacini}(1967)}]{pac67}
{Pacini}, F. 1967, \nat, 216, 567

\bibitem[{{Rea} \& {Esposito}(2011)}]{rea11}
{Rea}, N. \& {Esposito}, P. 2011, in High-Energy Emission from Pulsars and
  their Systems, ed. D.~F. {Torres} \& N.~{Rea}, 247

\bibitem[{{Rea} {et~al.}(2010){Rea}, {Esposito}, {Turolla}, {Israel}, {Zane},
  {Stella}, {Mereghetti}, {Tiengo}, {G{\"o}tz}, {G{\"o}{\u g}{\"u}{\c s}}, \&
  {Kouveliotou}}]{rea10}
{Rea}, N., {Esposito}, P., {Turolla}, R., {et~al.} 2010, Science, 330, 944

\bibitem[{{Str{\"u}der} {et~al.}(2001){Str{\"u}der}, {Briel}, {Dennerl},
  {Hartmann}, {Kendziorra}, {Meidinger}, {Pfeffermann}, {Reppin}, {Aschenbach},
  {Bornemann}, {Br{\"a}uninger}, {Burkert}, {Elender}, {Freyberg}, {Haberl},
  {Hartner}, {Heuschmann}, {Hippmann}, {Kastelic}, {Kemmer}, {Kettenring},
  {Kink}, {Krause}, {M{\"u}ller}, {Oppitz}, {Pietsch}, {Popp}, {Predehl},
  {Read}, {Stephan}, {St{\"o}tter}, {Tr{\"u}mper}, {Holl}, {Kemmer}, {Soltau},
  {St{\"o}tter}, {Weber}, {Weichert}, {von Zanthier}, {Carathanassis}, {Lutz},
  {Richter}, {Solc}, {B{\"o}ttcher}, {Kuster}, {Staubert}, {Abbey}, {Holland},
  {Turner}, {Balasini}, {Bignami}, {La Palombara}, {Villa}, {Buttler},
  {Gianini}, {Lain{\'e}}, {Lumb}, \& {Dhez}}]{str01}
{Str{\"u}der}, L., {Briel}, U., {Dennerl}, K., {et~al.} 2001, \aap, 365, L18

\bibitem[{{Sugizaki} {et~al.}(2001){Sugizaki}, {Mitsuda}, {Kaneda},
  {Matsuzaki}, {Yamauchi}, \& {Koyama}}]{sug01}
{Sugizaki}, M., {Mitsuda}, K., {Kaneda}, H., {et~al.} 2001, \apjs, 134, 77

\bibitem[{{Thompson} \& {Duncan}(1995)}]{tho95}
{Thompson}, C. \& {Duncan}, R.~C. 1995, \mnras, 275, 255

\bibitem[{{Thompson} {et~al.}(2002){Thompson}, {Lyutikov}, \&
  {Kulkarni}}]{tho02}
{Thompson}, C., {Lyutikov}, M., \& {Kulkarni}, S.~R. 2002, \apj, 574, 332

\bibitem[{{Tiengo} {et~al.}(2007){Tiengo}, {Mereghetti}, {Esposito}, {De Luca},
  \& {Gotz}}]{tie07atel}
{Tiengo}, A., {Mereghetti}, S., {Esposito}, P., {De Luca}, A., \& {Gotz}, D.
  2007, The Astronomer's Telegram, 1243, 1

\bibitem[{{Turner} {et~al.}(2001){Turner}, {Abbey}, {Arnaud}, {Balasini},
  {Barbera}, {Belsole}, {Bennie}, {Bernard}, {Bignami}, {Boer}, {Briel},
  {Butler}, {Cara}, {Chabaud}, {Cole}, {Collura}, {Conte}, {Cros}, {Denby},
  {Dhez}, {Di Coco}, {Dowson}, {Ferrando}, {Ghizzardi}, {Gianotti}, {Goodall},
  {Gretton}, {Griffiths}, {Hainaut}, {Hochedez}, {Holland}, {Jourdain},
  {Kendziorra}, {Lagostina}, {Laine}, {La Palombara}, {Lortholary}, {Lumb},
  {Marty}, {Molendi}, {Pigot}, {Poindron}, {Pounds}, {Reeves}, {Reppin},
  {Rothenflug}, {Salvetat}, {Sauvageot}, {Schmitt}, {Sembay}, {Short},
  {Spragg}, {Stephen}, {Str{\"u}der}, {Tiengo}, {Trifoglio}, {Tr{\"u}mper},
  {Vercellone}, {Vigroux}, {Villa}, {Ward}, {Whitehead}, \& {Zonca}}]{tur01}
{Turner}, M.~J.~L., {Abbey}, A., {Arnaud}, M., {et~al.} 2001, \aap, 365, L27

\bibitem[{{Ubertini} {et~al.}(2003){Ubertini}, {Lebrun}, {Di Cocco}, {Bazzano},
  {Bird}, {Broenstad}, {Goldwurm}, {La Rosa}, {Labanti}, {Laurent}, {Mirabel},
  {Quadrini}, {Ramsey}, {Reglero}, {Sabau}, {Sacco}, {Staubert}, {Vigroux},
  {Weisskopf}, \& {Zdziarski}}]{ube03}
{Ubertini}, P., {Lebrun}, F., {Di Cocco}, G., {et~al.} 2003, \aap, 411, L131

\bibitem[{{van der Horst} {et~al.}(2010){van der Horst}, {Connaughton},
  {Kouveliotou}, {G{\"o}{\u g}{\"u}{\c s}}, {Kaneko}, {Wachter}, {Briggs},
  {Granot}, {Ramirez-Ruiz}, {Woods}, {Aptekar}, {Barthelmy}, {Cummings},
  {Finger}, {Frederiks}, {Gehrels}, {Gelino}, {Gelino}, {Golenetskii},
  {Hurley}, {Krimm}, {Mazets}, {McEnery}, {Meegan}, {Oleynik}, {Palmer},
  {Pal'shin}, {Pe'er}, {Svinkin}, {Ulanov}, {van der Klis}, {von Kienlin},
  {Watts}, \& {Wilson-Hodge}}]{van10}
{van der Horst}, A.~J., {Connaughton}, V., {Kouveliotou}, C., {et~al.} 2010,
  \apjl, 711, L1

\bibitem[{{Vianello} {et~al.}(2009){Vianello}, {G{\"o}tz}, \&
  {Mereghetti}}]{via09}
{Vianello}, G., {G{\"o}tz}, D., \& {Mereghetti}, S. 2009, \aap, 495, 1005

\bibitem[{{Xu}(2007)}]{xu07}
{Xu}, R. 2007, Advances in Space Research, 40, 1453

\end{thebibliography}

\end{document}